\documentclass[journal]{IEEEtran}
\usepackage{cite}
\usepackage{amsmath,amssymb,amsfonts}
\usepackage{algorithmic}
\usepackage{graphicx}
\usepackage{textcomp}
\usepackage{balance}
\usepackage{array}
\newcolumntype{P}[1]{>{\centering\arraybackslash}p{#1}} % tables with specific column width and centered
\newcolumntype{M}[1]{>{\centering\arraybackslash}m{#1}} % tables with specific row width and centered
\usepackage{soul}
\usepackage{hyperref}
\usepackage{url}
\begin{document}
%%%%%%%%%%%%%%%%%%%%%%%%%%%%%%%%%%%%%%%%%%%%%%%%%%
\title{Radio Resource Management \\Techniques for Multibeam Satellite Systems}
%%%%%%%%%%%%%%%%%%%%%%%%%%%%%%%%%%%%%%%%%%%%%%%%%%
\author{Steven~Kisseleff,~\IEEEmembership{Member,~IEEE,}
        Eva~Lagunas,~\IEEEmembership{Senior~Member,~IEEE,}
		Tedros~Salih~Abdu,~\IEEEmembership{Student Member,~IEEE,}
        Symeon~Chatzinotas, ~\IEEEmembership{Senior~Member~IEEE}
        and Bj\"{o}rn Ottersten, \IEEEmembership{Fellow, IEEE.}% <-this % stops a space
\thanks{The authors are with the Interdisciplinary Centre for Security, Reliability and Trust (SnT), University of Luxembourg, L-1855 Luxembourg. E-mail: $\{$steven.kisseleff, eva.lagunas, tedros-salih.abdu, symeon.chatzinotas, bjorn.ottersten$\}$@uni.lu.}% <-this % stops a space
\thanks{The research leading to this publication has received funding from the Luxembourg National Research Fund (FNR) under the project FlexSAT (13696663), ROSETTA (11632107) and the bilateral FNR-ANR project SIERRA.}
%\thanks{Manuscript received on January 27, 2014.}% <-this stops a space
%\thanks{$^{*}$Corresponding author.}% <-this stops a space
}
%\markboth{IEEE Communication Letters}%
%{T.S. Abdu \MakeLowercase{\textit{et al.}}: Radio Resource Management Techniques for Multibeam Satellite Systems}
\maketitle
\begin{abstract}
Next–generation of satellite communication (SatCom) networks are expected to support extremely high data rates for a seamless integration into future large satellite-terrestrial networks. In view of the coming spectral limitations, the main challenge is to reduce the cost per bit, which can only be achieved by enhancing the spectral efficiency. In addition, the capability to quickly and flexibly assign radio resources according to the traffic demand distribution has become a must for future multibeam broadband satellite systems. This article presents the radio resource management problems encountered in the design of future broadband SatComs and provides a comprehensive overview of the available techniques to address such challenges. Firstly, we focus on the demand-matching formulation of the power and bandwidth assignment. Secondly, we present the scheduling design in practical multibeam satellite systems. Finally, a number of future challenges and the respective open research topics are described.
\end{abstract}
%%%%%%%%%%%%%%%%%%%%%%%%%%%%%%%%%%%%%%%%%%%%%%%%%%
%%%%%%%%%%%%%%%%%%%%%%%%%%%%%%%%%%%%%%%%%%%%%%%%%%
\section{Introduction}
\label{sec:Introduction}
%%%%%%%%%%%%%%%%%%%%%%%%%%%%%%%%%%%%%%%%%%%%%%%%%%
\IEEEPARstart{S}{atellite} manufacturers and satellite operators are working hard to launch the next generation of flexible Ultra High Throughput Satellite (UHTS) systems able to offer Terabit per second in–orbit capacity when and where needed. These future satellite architectures are expected to unlock new applications with cost–effective ubiquitous connectivity and provide a competitive alternative to the data–hungry digital society \cite{symo2018}. Novel services are expected to generate time-varying and non-uniformly distributed traffic demand, thus leading to even larger fluctuations of this heterogeneous demand. Accordingly, the required throughput enhancements can only be achieved by pushing forward the multibeam architecture with reduced beam size, taking advantage of frequency reuse and adaptation of the satellite capacity  \cite{Oltjon_2020}. The main objective for the design of the next generation of UHTS is three–fold:
\begin{itemize}
    \item Decrease the cost of satellite services by improving the satellite capacity through enhancement of the spectral efficiency;
    \item Increase the flexibility for the satellite service provisioning by dynamically allocating the available radio resources according to the time-varying traffic demands;
    \item Improve the connectivity in the under-served areas using non-geostationary orbit (NGSO) satellites in the low (LEO) and medium (MEO) orbits in addition to the well established geostationary (GEO) satellites.
\end{itemize}

Targeting the aforementioned goals, extensive research has been conducted in the past years in order to establish systematic design methodologies that aim at exploiting all available degrees of freedom for the optimization of radio resources. 

In this work, we attempt to summarize these design guidelines. Specifically, we discuss the key performance metrics and fundamental radio resource optimization problems and techniques encountered when dealing with multibeam UHTS systems. In addition, we address some of the most promising advancements as well as remaining research challenges.

This paper is organized as follows. In Section \ref{sec:RRO}, we discuss the radio resource management (RRM) methods. In particular, we consider the optimization in time, frequency and spatial domains as well as power allocation. In Section \ref{sec:opencha}, we provide our vision on the future directions and open challenges for the UHTS. Finally, the paper is concluded in Section \ref{sec:conc}.

%%%%%%%%%%%%%%%%%%%%%%%%%%%%%%%%%%%%%%%%%%%%%%%%%%
\section{Radio Resource Management}
\label{sec:RRO}
%%%%%%%%%%%%%%%%%%%%%%%%%%%%%%%%%%%%%%%%%%%%%%%%%%

We consider the downlink transmissions from the satellite providing broadband services to a geographical area by means of $L$ beams. Within the coverage area of each beam $l$, $N_l$ single antenna users are randomly scattered. Their aggregate traffic demand is denoted as $D_l$. To satisfy the demand of each user in each beam, the satellite may employ any of the traditional multi-user access techniques, typically time- or frequency-division multiplexing, i.e. TDM or FDM, respectively. Furthermore, spatial multiplexing can be employed, which however does not guarantee full orthogonality of the adjacent data streams. Here, spatial multiplexing\footnote{Since beam patterns are more difficult to reconfigure and are correspondingly much less flexible than the precoding, we assume a fixed beam pattern and focus on spatial multiplexing in terms of precoding in this work.} may refer to the multiple beams or to precoding using multiple transmit antennas at the satellite. Hence, the most common RRM problems target the optimal distribution of signal bandwidth, time slots and transmit power among the users as well as the design of precoding vectors \cite{Cocco_2018}, \cite{bhavani}. Note that neighboring beams may produce interference for the adjacent users, which needs to be accounted for in the system design.

A common strategy for addressing such problems is to consider a single representative super-user per beam with the assigned aggregate demand $D_l$ of all users to be served by that beam. This allows for the substantial complexity reduction without any change in system performance. Hence, the power and carrier allocation per beam can be decoupled from the design of scheduling\footnote{Note that beam hopping is a special case of scheduling and will be covered in Section \ref{sec:illuminate}.} and precoding per user. In the following, we take a closer look at these two parts of the RRM.

%%%%%%%%%%%%%%%%%%%%%%%%%%%%%%%%%%%%%%%%%%%%%%%%%%%%%%
\subsection{Bandwidth and power allocation}
\label{sec:Band_Pow}
%%%%%%%%%%%%%%%%%%%%%%%%%%%%%%%%%%%%%%%%%%%%%%%%%%%%%%
For the systematic design of user modems, the bandwidth allocation is typically done via assignment of carriers with equal width. Hence, the number of assigned carriers and their position within the given frequency range are optimized to achieve good signal quality with the minimum resources.

For this, we assume that the whole frequency band of total width $B_{\mathrm{total}}$ is split among $K$ carriers of equal width $B_c=B_{\mathrm{total}}/K$. 
Often, the carrier assignment problem is solved suboptimally via orthogonal splitting of the resource, which leads to the so-called frequency reuse. This solution is characterized by the number of carriers (colors) to be assigned to the beams in such a way that the distance between the beams with the same color is maximized. However, the drawback of this solution is the strict orthogonality of the frequency bands, which is not always required and limits the spectral efficiency. 

For a better spectrum utilization, a more thorough RRM is needed, which accounts for the inter-beam interference, cf. \cite{park2012,tedros}. The resource optimization variables are:
\begin{itemize}
    \item $x_{k}[l]\in \{0,1\}$ denotes the assignment vector for carrier $k$, such that $x_{k}[l]=1$ indicates that carrier $k$ is assigned to the $l$th beam;
    \item $p_{k}[l]$ is the transmit power of $k$th carrier in the $l$th beam.
\end{itemize}

We denote the channel power gain for the transmission of the $m$th beam received by the super-user of the $l$th beam as $g_{l}[m]$. Note that this channel gain is assumed to be frequency-flat due to the typical line-of-sight (LoS) propagation in SatComs and fixed rain attenuation margin. Hence, we can formulate the signal-to-interference-plus-noise ratio (SINR) for each super-user in carrier $k$ as $\gamma_{l,k}=\frac{g_{l}[l] p_{k}[l]x_{k}[l] } {\sum_{m\neq l} g_{l}[m]p_{k}[m]x_{k}[m] + \sigma_l^2}$,
%\begin{equation}
%\label{eq:1}
%    \gamma_{l,k}=\frac{g_{l}[l] p_{k}[l]x_{k}[l] } {\sum_{m\neq l} g_{l}[m]p_{k}[m]x_{k}[m] + \sigma_l^2},
%\end{equation}
where $\sigma_n^2$ is the power of the received additive white Gaussian noise (AWGN). Hence, the offered capacity in beam $l$ is given by $C_{l}=\sum^{K}_{k=1}  B_{c} \log_2 (1 + \gamma_{l,k})$.
%\begin{equation}
%\label{eq:2}
%   C_{l}=\sum^{K}_{k=1}  B_{c} \log_2 (1 + \gamma_{l,k}).
%\end{equation}

For the traffic matching, the offered capacity $C_{l}$ should be as close as possible to the traffic demand $D_l$. There are different ways to formulate such objective mathematically. Some works make use of the Minimum Mean Squared Error (MMSE), which can be written as $\frac{1}{L}\sum^{L}_{l=1}\left(C_l-D_l\right)^2$.
Although this objective function is very useful for the satisfaction of the demand on average, it aims at minimizing both the unmet and unused capacities in the same way although the unused capacity is of course much less harmful. Another option is to maximize the minimum relative demand satisfaction. The corresponding objective function is written as $\min_l\{\frac{C_l}{D_l}\}$. This objective targets the largest relative unmet capacity, i.e. $\max_l\{1-\frac{C_l}{D_l}\}$. Correspondingly, this objective is more useful for the service provision than the MMSE objective. However, only the beam with the lowest demand satisfaction is considered while the satisfaction of other beams is neglected, which is a drawback of this objective. Instead, the Unmet System Capacity (USC) has been increasingly employed as objective function, which can be expressed as $\sum_{l=1}^L\min\{C_l-D_l,0\}$. This objective has the advantage of targeting the unmet capacity in all beams.

The available radio resources are taken into account in the optimization in terms of constraints, which may depend on the payload architecture. As an example, multiple beams can be served by a single transponder, which implies the restriction of the resources per transponder.

An example of the optimization problem for the traffic matching with classical (linear) constraints is given below: 
\begin{eqnarray}
\label{carrier_power_prob}
&&\hspace*{-8mm}\operatorname*{\mathrm{maximize}}_{p_k[l], x_k[l],\:\forall k, l}\sum_{l=1}^L\min\{C_l-D_l,0\},\\
%&&\hspace*{-8mm}\mathrm{s.t.:}\ \mathrm{C1)}\: C_l=\sum^{K}_{k=1}  B_{c} \log_2 (1 + \frac{g_{l}[l] p_{k}[l]x_{k}[l] } {\sum_{m\neq l} g_{l}[m]p_{k}[m]x_{k}[m] + \sigma_l^2}),\notag\\
&&\hspace*{-8mm}\mathrm{s.t.:}\ \mathrm{C1)}\: \sum_{k,l}p_k[l]\leq P_{\mathrm{total}},\ \mathrm{C2)}\: p_k[l]\geq 0,\notag\\
&&\mathrm{C3)}\: x_k[l]\in\{0, 1\},\:\forall k, l.\notag
\end{eqnarray}
Such optimization problems are in most of the cases non-linear and non-convex due to the logarithm function as well as non-linear dependencies of the SINR on the optimization parameters. Furthermore, the binary carrier assignment indicator $x_k[l]$ makes it a mixed-integer program. Hence, no optimal solution can be determined using the known methods of convex optimization, i.e. \cite{Boyd2004}. Instead, suboptimal methods are proposed, which tackle parts of the problem separately and then iteratively tune the parameters. The splitting of the problem is done in such a way that power allocation and carrier assignment are separated \cite{Lei2010}. In order to solve the non-convex problems, the successive convex approximation technique can be applied. Specifically, the non-convex part of the objective function or of a constraint is identified and set to a constant value. Through this, the problem becomes convex and can be optimally solved. In the next iteration, the value of the non-convex part is updated and the procedure repeats itself until convergence. The dependencies related to integer or binary parameters, i.e. $x_k[l]\:\forall k, l$ can be tackled separately as well. This part of the optimization is then solved either via combinatorial methods, such as e.g. Hungarian algorithm, or via approximation. In the latter case, integer values are replaced by continuous values, e.g. $0\leq x_k[l]\leq 1$ instead $x_k[l]\in\{0, 1\}$.

Alternatively, metaheuristics and machine learning based methods can be applied in order to determine the optimal carrier and power allocation \cite{Paris2019,lei_lagunas}. In this case, the features associated with the power or carrier allocation can be learnt and predicted using only the traffic demand $D_l$ as input.

%%%%%%%%%%%%%%%%%%%%%%%%%%%%%%%%%%%%%%%%%%%%%%%%%%%%%%
\subsection{Scheduling}
%%%%%%%%%%%%%%%%%%%%%%%%%%%%%%%%%%%%%%%%%%%%%%%%%%%%%%
The RRM in time domain is referred to as scheduling and usually corresponds to the user selection and aggregation of the information to be transmitted in a physical layer frame. In this context, the combination of precoding with user scheduling imposes additional challenges to the satellite RRM design. Furthermore, beam scheduling has recently emerged in the context of Beam Hopping (BH), where a proper beam illumination pattern needs to be designed. Below we discuss these topics in more detail.

%%% EVA please complete these two parts based on CADSAT and FlexPreDem. Do we need both subsections?
% \subsubsection{Scheduling for carrier aggregation}
% (Load balancing)

%%%%%%%%%%%%%%%%%%%%%%%%%%%%%%%%%%%%%%%%%%%%%%%%%%%%%%
\subsubsection{User scheduling}
\label{sec:Sched}
%%%%%%%%%%%%%%%%%%%%%%%%%%%%%%%%%%%%%%%%%%%%%%%%%%%%%%
User scheduling refers to the selection and aggregation of
the information to be transmitted in each PHY frame. Since 2003, Adaptive Code and Modulation (ACM) scheme is applied, assigning to each PHY frame the most suitable modulation and code (ModCod) value according to the measured SINR. In the DVB-S2(X) standard, however, the choice of the physical layer mode to be used
in each frame is necessarily linked to the scheduling process,
as all the user packets included in one frame are transmitted
with the same ModCod. Clearly, the larger the difference
among the users’ SINR encoded within a frame, the higher the
performance loss. This motivates a scheduling design based on the  users SINR levels similarity.

%%%%%%%%%%%%%%%%%%%%%%%%%%%%%%%%%%%%%%%%%%%%%%%%%%%%%%
\subsubsection{User scheduling and precoding}
\label{sec:Sched_Prec}
%%%%%%%%%%%%%%%%%%%%%%%%%%%%%%%%%%%%%%%%%%%%%%%%%%%%%%
%Given the available signal bandwidth $B_l=B_c\sum_{k}x_k[l]$ and allocated power per beam $P_l=\sum_kp_k[l]$, the management of radio resource with respect to the individual demand of the scattered users can be addressed. 
%In this context, 
Joint scheduling and precoding design has been thoroughly investigated for terrestrial and SatComs in the past years, cf. \cite{7811843}. This is because both aspects are closely related and heavily impact each others' performance.%the user scheduling has an important impact on the final precoding performance. 

The precoding design refers to the design of a set of precoding vectors $\textbf{W}=\begin{bmatrix} \textbf{w}_1 & \textbf{w}_2 & \cdots & \textbf{w}_L \end{bmatrix}$, one for each of the users to be simultaneously served. Note that we assume a single user per beam to be served during a given time slot. 

The received signal vector can be expressed as $\textbf{y}=\textbf{H}\textbf{x}+\textbf{n}$, with $\textbf{H}=\begin{bmatrix} \textbf{h}_1 & \textbf{h}_2 & \cdots & \textbf{h}_L \end{bmatrix}^H$, $\textbf{y}=\begin{bmatrix} y_1 & y_2 & \cdots & y_L \end{bmatrix}^T$ and $\textbf{n}=\begin{bmatrix} n_1 & n_2 & \cdots & n_L \end{bmatrix}^T$.

The channel vectors $\textbf{h}_l \in \mathbb{C}^{L \times 1}$ describe the channel from the satellite to the user located in the $l$-th beam.

A widely used precoding design in the satellite community follows the Regularized Zero-Forcing (ZF) form, which makes use of the so-called regularized pseudo-inverse, cf. \cite{406634},
\begin{equation}
	\textbf{W}_{RZF}=\eta^{\prime}\textbf{H}^H\left(\textbf{H}\textbf{H}^H + \alpha\textbf{I}\right)^{-1},
	\label{eq:rzf}
\end{equation}
where $\alpha > 0$ is the regularization factor, and $\eta^{\prime}=\sqrt{P_{\text{total}}/\text{Trace}\left\{\textbf{W}_{RZF}\textbf{W}^H_{RZF}\right\}}$.

The main idea of \eqref{eq:rzf} is to enforce $\textbf{h}^{H}_{n}\textbf{w}_i = 0$, for $n\neq i$. However, depending on the structure of $\textbf{H}$, which is determined by the location of the scheduled user, the inversion becomes challenging. To overcome this issue, user scheduling design is mostly focused on grouping users with channel vectors as orthogonal as possible. This is essentially the basis of the semi-orthogonality criteria originally proposed in \cite{yoo}.

Joint precoding and scheduling design is a challenging problem, particularly when combined with the requirements dictated by the DVB standard described in section \ref{sec:Sched}. The joint design is a coupled problem. Initially, the users are scheduled based on similar link/SINR conditions. Then, the precoding is designed for the selected users, which in turn affects the users' perceived SINR \cite{Ashok2019}.

%%%%%%%%%%%%%%%%%%%%%%%%%%%%%%%%%%%%%%%%%%%%%%%%%%%%%%
\subsubsection{Beam illumination design}
\label{sec:illuminate}
%%%%%%%%%%%%%%%%%%%%%%%%%%%%%%%%%%%%%%%%%%%%%%%%%%%%%%
In BH systems, all  the  available satellite resources are employed to provide service to a certain subset  of  beams,  which  is  active  for  some  portion  of  time, dwelling  just  long  enough  to  fill  the  demand  in  each  beam. The set of illuminated beams changes in each time-slot based on  a  time-space  transmission  pattern that is periodically repeated. Typically, all bandwidth is employed for each illuminated beam, which are spatially isolated ensuring that the co-channel interference is kept to minimal. The main design problem in BH consists in finding the optimal beam illumination pattern, i.e. which beams are activated when and for how long. Periodic beam illumination pattern is preferred for synchronization aspects, as the terminal on/off switching occurs at predictable times \cite{freedman}.

The formulation of the beam scheduling considers a specific time window $T_w$ segmented into time-slots of duration $T_s$ (which represents the minimum slot allocation unit). Within a BH window $T_w$, a number of illuminated snapshots are used. Herein, we use ``snapshot'' to refer to a particular arrangement of
illuminated and un-illuminated beams. By defining $\mathcal{G}$ as the set of possible snapshots, we can formulate a fairness-oriented beam illumination design problem as
\begin{eqnarray}
\label{bh_prob}
&&\hspace*{-8mm}\operatorname*{\mathrm{maximize}}_{t_1,\ldots,t_G}\ \min_l\left\{ \frac{C_l}{D_l} \right\},\\
%&&\hspace*{-8mm}\mathrm{s.t.:}\ \mathrm{C1)}\: C_l=\sum^{K}_{k=1}  B_{c} \log_2 (1 + \frac{g_{l}[l] p_{k}[l]x_{k}[l] } {\sum_{m\neq l} g_{l}[m]p_{k}[m]x_{k}[m] + \sigma_l^2}),\notag\\
&&\hspace*{-8mm}\mathrm{s.t.:}\  \mathrm{C1)}\ \sum_{g\in\mathcal{G}}t_g T_{s}=T_H,\notag\\
&& \mathrm{C2)}\ \{ t_1,\ldots,t_G \} \ \text{integer}.\notag
\end{eqnarray}
The optimization variables in \eqref{bh_prob} are the integers $t_1,\ldots,t_G$, and as a consequence,  \eqref{bh_prob} is a mixed integer linear programming
problem (MILP). To simplify the proposed problem, a reformulation relaxing the integers to non-negative continuous numbers is usually pursued. Next, the objective function is transformed from a max-min to a maximization-only by introducing an auxiliary variable $\eta=\min_l \{\frac{C_l}{D_l}\}$ or, in other words, $\frac{C_l}{D_l}\geq \eta$, $l=1,\ldots,L$.

%The  time  slot  assignment  matrix  can  be  formulated  as a binary matrix $\boldsymbol\Psi$ of dimension $( N_t \times L )$. With this definition, an SINR expression can be obtained for each time-slot $t$ as $\gamma_{l,k,t}=\frac{g_{l}[l] p_{k}[l]x_{k}[l] \psi_{t,l} } {\sum_{m\neq l} g_{l}[m]p_{k}[m]x_{k}[m]\psi_{t,m} + \sigma_l^2}$. Note that the carrier assignment variables is intentionally kept in the notation, since the beam hopping pattern can be optimized together with the spectrum assignment.
%\begin{equation}
%\label{eq:sinr_BH}
%    \gamma_{l,k,t}=\frac{g_{l}[l] p_{k}[l]x_{k}[l] \psi_{t,l} } {\sum_{m\neq l} g_{l}[m]p_{k}[m]x_{k}[m]\psi_{t,m} + \sigma_l^2}.
%\end{equation}

In most of the cases, the illumination time per beam is proportional to the respective demand. Further constraints to be considered in the BH design are related to: (i) number of simultaneously active beams per time slot, (ii) switching frequency (if too high can negatively affect the terminal stability and overall efficiency) (iii) latency perceived at the user side. Once the illumination pattern is fixed, optimization of the same key performance indicators as the ones in Section \ref{sec:Band_Pow} can be pursued.

\section{Open Challenges}
\label{sec:opencha}
%%%%%%%%%%%%%%%%%%%%%%%%%%%%%%%%%%%%%%%%%%%%%%%%%%
In this section, we will address some of the open research challenges for the RRM in SatComs.
%%%%%%%%%%%%%%%%%%%%%%%%%%%%%%%%%%%%%%%%%%%%%%%%%%
%%%%%%%%%%%%%%%%%%%%%%%%%%%%%%%%%%%%%%%%%%%%%%%%%%
\subsection{Reconfigurable and steerable beams}
%%%%%%%%%%%%%%%%%%%%%%%%%%%%%%%%%%%%%%%%%%%%%%%%%%
The most common design of multibeam satellite systems is based on a regular beam pattern with equal beam width for all beams. Such a regular pattern is very beneficial, since it reduces the complexity of the optimization. As an example, a simple frequency reuse scheme can be applied, which minimizes the inter-beam interference. On the other hand, some of the beams may cover very crowded areas with extremely high aggregated demand. Hence, narrow beams with a high antenna gain may be preferable in such cases while large areas with a low population density can be covered by a wide beam. Hence, irregular beam patterns seem promising. 

\begin{figure}
    \centering
    \includegraphics[width=0.25\textwidth]{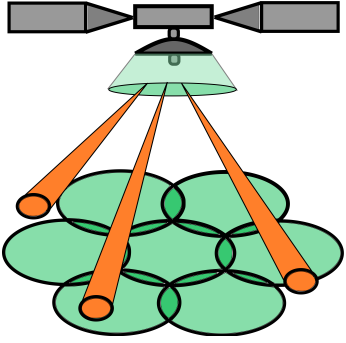}
    \caption{Joint application of active and passive antennas. Green beams employ traditional passive antenna technique to provide a basic coverage for areas with low-to-moderate demand. Orange beams employ active antennas to track high-demand areas.}
    \label{fig_multibeam}
\end{figure}

Active steerable narrow-beam antennas are known to be very effective in tracking of the high-demand areas (cf. \cite{893245}) while passive antennas would provide a basic coverage for all other areas within the field of view, see Fig. \ref{fig_multibeam}. 
Correspondingly, the time/frequency plan for the steerable beams needs to be jointly optimized in order to avoid uncontrolled interference among the beams. However, such a joint architecture with a reconfigurable beam pattern and multiple active antennas is very challenging, especially from the hardware perspective.

Reconfigurable beams are especially useful for smaller NGSO satellites, which may reconfigure the beams ``on the fly'', since the demand distribution within the coverage area of these satellites frequently changes. However, the mobility of these NGSO satellites poses additional challenges with respect to the synchronization and channel acquisition as well as accurate estimation of the traffic demand. The resulting uncertainties may lower the overall system performance. 

In general, reconfigurable beams introduce a few additional unknowns, such as beam width, shape, and orientation. These parameters would make the RRM problems more challenging and less tractable analytically. Hence, we can anticipate the increased use of machine learning methods in future.
%\subsection{Non-orthogonal multiple access}
%Non-orthogonal multiple access (NOMA) is one of the key techniques, which can help to dramatically improve the throughput of broadband transmissions. The idea is to transmit multiple data streams over the same communication channel. For this, the data streams need to be sufficiently separable, e.g. in terms of receive power, such that the maximum number of supported data streams is limited to a single digit. But the corresponding throughput increase is still substantial. The separation of the streams at the receivers is done via successive interference cancellation (SIC). In the context of SatComs, NOMA can be applied in order to satisfy a very high demand, which would otherwise relate to high power or bandwidth consumption at the satellite, cf. \cite{8904429}. Here, multiple users belonging either to the same beam but separated in space or to the neighboring beams can be aligned in a single transmission utilizing the NOMA technique and SIC.

%For the RRM, such an extension poses a big challenge, since the distribution of the channel gains of the individual users may need to be taken into account in the carrier assignment and power allocation as well as scheduling and precoding, such that a joint optimization of all these components is difficult to avoid.
%%%%%%%%%%%%%%%%%%%%%%%%%%%%%%%%%%%%%%%%%%%%%%%%%%
\subsection{Hardware impairments}
%%%%%%%%%%%%%%%%%%%%%%%%%%%%%%%%%%%%%%%%%%%%%%%%%%
Hardware impairments can affect the accuracy of the designed RRM solution in various ways. Among the most harmful impairments for the downlink transmissions, the non-linearity of power amplifiers poses a real challenge for the system design. The non-linear distortion due to the operation in the non-linear region of the amplifier manifests itself as power leakage from the distorted carrier into the adjacent carriers. The leakage generates additional interference, which depends on the signal magnitude and the distance between carriers. To solve this issue, either pre-distortion methods are applied or the frequency plan is adjusted in order to account for this effect. The former leads to a substantial increase in hardware complexity, cf. \cite{7227095}. The latter introduces additional constraints for the optimization problem, thus making it even more difficult to tackle in real-time.

%%%%%%%%%%%%%%%%%%%%%%%%%%%%%%%%%%%%%%%%%%%%%%%%%%
%\subsection{Resource management for low-orbit constellations}
%%%%%%%%%%%%%%%%%%%%%%%%%%%%%%%%%%%%%%%%%%%%%%%%%%
%Advances in satellite design, manufacturing and launch service capabilities have enabled the deployment of NGSO constellations. However, NGSO satellites are characterized by rapidly moving orbits which require an extra effort regarding the satellite resource management.

%Apart from mobility, each NGSO satellite operates multiple dynamic individual narrow beams, whose pointing and time/frequency plan has to be carefully designed to satisfy the users' quality of service while optimizing the satellite resources according to the system requirements.

%%%%%%%%%%%%%%%%%%%%%%%%%%%%%%%%%%%%%%%%%%%%%%%%%%
\subsection{Low complexity optimization}
%%%%%%%%%%%%%%%%%%%%%%%%%%%%%%%%%%%%%%%%%%%%%%%%%%
One of the main challenges for the future SatComs is the complexity of the optimization algorithms. In fact, the dynamics of the demand may increase in future together with its average, which requires a frequent recalculation of the resource allocation. This is especially crucial for the NGSO satellites, e.g. LEO and MEO satellites, which have to quickly adapt to the changes in the field of view. Hence, the complexity of the algorithms needs to be sufficiently low in order to enable their timely application.

Furthermore, since the cooperation of multiple satellites is envisioned for a seamless operation, their coverage areas may overlap, which requires additional efforts to avoid not only the inter-beam, but also the inter-satellite interference. This puts even stricter requirements on the computational complexity, because the joint coverage area and the number of variables increase dramatically. In this case, the federated learning methods seem promising. 

%%%%%%%%%%%%%%%%%%%%%%%%%%%%%%%%%%%%%%%%%%%%%%%%%%
\subsection{Increased flexibility with multiple degrees of freedom}
%%%%%%%%%%%%%%%%%%%%%%%%%%%%%%%%%%%%%%%%%%%%%%%%%%
Active antennas and on-board digital processors are two technological enablers that have opened a door to advanced RRM strategies for flexible satellite systems. This concept has many degrees of freedom (power, carriers, bandwidth, beam pattern). However, it is not entirely clear, whether all of them are needed. Since each type of flexibility comes with an additional layer of complexity which could eventually be a point of failure, the relative gain provided by each individual level of flexibility needs to be carefully assessed.

Furthermore, it is not evident what degree of flexibility is required by the actual traffic demand distribution. Therefore, a connection between the different Service Level Agreements (SLAs) existing in modern broadband system and the degree of flexibility required to match these SLA should be investigated. 

%%%%%%%%%%%%%%%%%%%%%%%%%%%%%%%%%%%%%%%%%%%%%%%%%%
\subsection{Resource allocation for system orchestration}
%%%%%%%%%%%%%%%%%%%%%%%%%%%%%%%%%%%%%%%%%%%%%%%%%%
Depending on the type of satellites, i.e. GEO, MEO or LEO, the SatCom systems are typically treated differently according to the features of the respective orbits, e.g. high latency for GEO satellites or high Doppler shift for LEO satellites. Correspondingly, the design approach for each type of satellite constellation is different, which will pose a burden for the orchestration of the network in future, since it leads to compatibility problems both for the ground segment network operations (NOC) and space segment payload operations (SOC). The interplay between NOC and SOC will certainly become very challenging with increasing number of satellites as well as traffic demand and stricter service requirements without a proper management of connectivity and resource allocation for satellites belonging to different orbits \cite{shree}. Instead, the envisioned joint operation of multiple constellations would enable the intelligent network-oriented RRM in future SatComs and would lead to the cloudification of the system.

\begin{figure}
    \centering
    \includegraphics[width=0.48\textwidth]{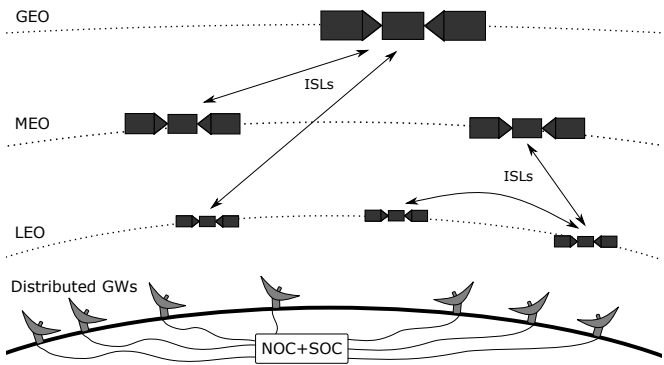}
    \caption{Joint operation of GEO, MEO and LEO satellites in a multi-constellation architecture. The space segment is supported by ISLs between the satellites of the same and different constellations. Together with a set of distributed GWs the ISLs enable the cloudification of SatComs.
    }
    \label{fig_constellations}
\end{figure}

In order to enable such a multi-constellation SatCom network, a suitable dynamic topology needs to be designed, which may go beyond the traditional star and mesh. This novel technology will be supported by intersatellite links (ISLs) and distributed gateways (GWs), see Fig. \ref{fig_constellations}. 
Furthermore, network slicing seems promising and, combined with a holistic resource allocation, would make the system orchestration more intelligent and adaptive. However, a joint design of a multi-constellation system is extremely challenging, since it accumulates all challenges associated with the design of the respective individual constellations.

%%%%%%%%%%%%%%%%%%%%%%%%%%%%%%%%%%%%%%%%%%%%%%%%%%
\section{Conclusion}
\label{sec:conc}
%%%%%%%%%%%%%%%%%%%%%%%%%%%%%%%%%%%%%%%%%%%%%%%%%%
In this paper, the fundamental optimization strategies associated with radio resource management for the multibeam SatComs have been presented. The hierarchical optimization comprises bandwidth and power allocation among the beams followed by scheduling and precoding. The respective optimization problems and design techniques have been explained. In addition, open challenges and trends for the multibeam satellite systems have been discussed. In this context, future research directions have been motivated.
% \section*{Acknowledgment}

%\vfill
\bibliographystyle{IEEEtran}
\bibliography{IEEEabrv,refs_COMML}
%%%%%%%%%%%%%%%%%%%%%%%%%%%%%%%%%%%%%%%%%%%%%%%%%%
%%%%%%%%%%%%%%%%%%%%%%%%%%%%%%%%%%%%%%%%%%%%%%%%%%
% biography section HERE
%\end{IEEEbiography}
%%%%%%%%%%%%%%%%%%%%%%%%%%%%%%%%%%%%%%%%%%%%%%%%%%
%%%%%%%%%%%%%%%%%%%%%%%%%%%%%%%%%%%%%%%%%%%%%%%%%%
%%%%%%%%%%%%%%%%%%%%%%%%%%%%%%%%%%%%%%%%%%%%%%%%%%
%%%%%%%%%%%%%%%%%%%%%%%%%%%%%%%%%%%%%%%%%%%%%%%%%%
%%%%%%%%%%%%%%%%%%%%%%%%%%%%%%%%%%%%%%%%%%%%%%%%%%
%%%%%%%%%%%%%%%%%%%%%%%%%%%%%%%%%%%%%%%%%%%%%%%%%%
\end{document}